\documentclass[intlimits,twoside,a4paper]{article}
\usepackage{graphicx}
\usepackage[T2A]{fontenc}
\usepackage[cp1251]{inputenc}
\usepackage[eqsecnum]{cmpj3}

\issue{2023}{26}{1}{13801}
\doinumber{10.5488/CMP.26.13801}

\title[On the role of a conference in life of scientist]{One for all and all for one: on the role of a conference in a scientist's life}

\author[O. Mryglod]{O. Mryglod\orcid{0000-0003-4415-7061}}

\address{Institute for Condensed Matter Physics of the National Academy of Sciences of Ukraine,\\
	1 Svientsitskii St., 79011 Lviv, Ukraine; $\mathbb{L}^4$ Collaboration \& Doctoral College for the Statistical Physics of Complex Systems,
	Leipzig-Lorraine-Lviv-Coventry, Europe}


\date{Received December 4, 2022, in final form December 14, 2022}

\begin{document}
\maketitle
\begin{abstract}
The quantitative description of the scientific conference MECO (Middle European Cooperation in Statistical Phy\-sics) based on bibliographic records is presented in the paper. Statistics of contributions and participants, co-authorship patterns at the levels of authors and countries, typical proportions of newcomers and permanent participants as well as other characteristics of the scientific event are discussed. The results of this case study contribute to better understanding of the ways of formalization and assessment of conferences and their role in individual academic careers. To highlight the latter, the change of perspective is used: in addition to the general analysis of the conference data, an ego-centric approach is used to emphasize the role of a particular participant for the conference and, vice versa, the role of MECO in the researcher's professional life. This paper is  part of the special CMP issue dedicated to the anniversary of Bertrand Berche --- a well-known physicist, an active member of the community of authors and editors of the journal, long time collaborator and dear friend of the author.
\printkeywords
%
\end{abstract}

\section{Introduction}

\noindent \emph{Four hundred years ago, Francis Bacon said:\\ ``Reading
maketh a full man; conference a ready man,\\ and writing
an exact man'' \cite{Scott1908}.}\\

Today, we all are used to hearing how important  are the relations between researchers. The vitality of information exchange for the research process comes out even from Newton's famous statement ``If I have seen further, it is by standing on the shoulders of Giants'': the researcher is not a lone seeker of truth, but one of many actors united by the collectively gathered knowledge. Processing of the research results and, thus, the involvement in communication activities are  integral parts of a researcher's work. Ways of exchanging information evolve with time and development of technologies. Private discussions in small communities of ancient thinkers gave way to academic correspondence and now to various kinds of communication through different social media. However, even in the times of virtual relationships direct social interactions play a crucial role \cite{Hoekman2010}. Therefore, the regular scientific meetings are part of normal professional life of any researcher. Among other important features of scientific conferences, the establishment and fortification of personal interlinks is perhaps the most fundamental. Live communication provides an opportunity to exchange not only by ready-made ideas, but also by thoughts and reflections which lead to new answers and, sometimes even more importantly, to new questions \cite{King1961}.  Participation in scientific conferences is considered as one of the indicators of implementation of new results, at least at the level of integration into the existing knowledge used by colleagues. Therefore, it is difficult to overestimate the role of professional gatherings of academics.

Like many other human activities, scientific conferences can be characterized by various `digital footprints': bibliographic data related to conference publications, attendance statistics, information about conference organizers and venues --- all of which can be analysed to gain insight about disciplinary spectrum, key actors, collaboration patterns, geographic landscape, potential gender effects, editorial policies and many other aspects related to a specific event or scientific conferences in general, see, e.g.,~\cite{Fathalla2017,Barbosa2017,Larson2020,Else2019,Oswald2020,Bartneck2017,Martins2010,Souto2007}.
Still, it is also interesting to ask the research question in reverse, from the perspective of a country, an institution or individual researcher: what is the role of a particular conference in gaining a publication impact, in establishing collaborative links or in selecting research topics?

This paper contains selected results of analysis of the data related to the conference ``Middle European Cooperation in Statistical Physics'' (MECO) \footnote{See \url{https://sites.google.com/site/mecoconferencephysics/home} and \url{https://en.wikipedia.org/wiki/Middle_European_Cooperation_in_Statistical_Physics}}. Another ``point of attraction'' of interest here is determined by the main idea of the current CMP issue, i.e., the anniversary of a well known French physicist and a good friend of the author, Prof. \href{https://orcid.org/0000-0002-4254-807X}{Bertrand Berche}. He is an important part of the MECO community: on the one hand, he is among the people who shape this event as a member of the International Advisory Board since 2001, and on the other hand, he has personally participated in 21 conferences and belongs to TOP15 most contributive authors (with 31 papers). In other words, MECO conference is a significant part of Bertrand's academic career, which dates back to his first publication in 1989; and Bertrand, in his turn, is an important actor in 47-years-length MECO history. One can make such conclusions intuitively, but they are fortified by quantitative results further in the paper.

The structure of the paper is as follows: general information about the MECO conference, accompanied by the	results of a quantitative analysis of relevant bibliographic data, is presented in section~\ref{aboutMECO}; speculation about the role of Prof.~Berche in the history of MECO and the hypothetical impact of the MECO conference on Bertrand's academic career can be found in section~\ref{BB}; the concluding remarks are given in the last section~\ref{summary}.

\section{MECO conferences}\label{aboutMECO}
The Conference of the Middle European Cooperation in Statistical Physics was born from the idea to bridge the gap between the Eastern and Western European scientific community divided by the Iron Curtain \cite{Folk2021}. From the very beginning, special attention was paid to the geography of participants: it was decided to change the location every year and, in addition, to ensure that the organizations hosting the conference were chosen each time on the other side of the invisible line that separated one part of Europe from another \cite{Folk2021}. Due to these political-geographical aspects, the MECO conference can be considered as a symbolic bridge for scientific contacts not only at the personal level, but also at the level of countries. This conference is also a valuable object for scientometric study, as its history is already 48 years long: 47 meetings have been held since 1974.
Due to some data gaps, further quantitative analysis is done for 42 MECO events\footnote{No sources found for data on the 14th meeting in 1987, only partial data can be collected for the 6th (1979), 35th (2010), 38 (2013) and 47th (2022) events.}: a set of bibliographic records compiled for 3752 oral and poster contributions.

\begin{figure}
\includegraphics[width=0.48\textwidth]{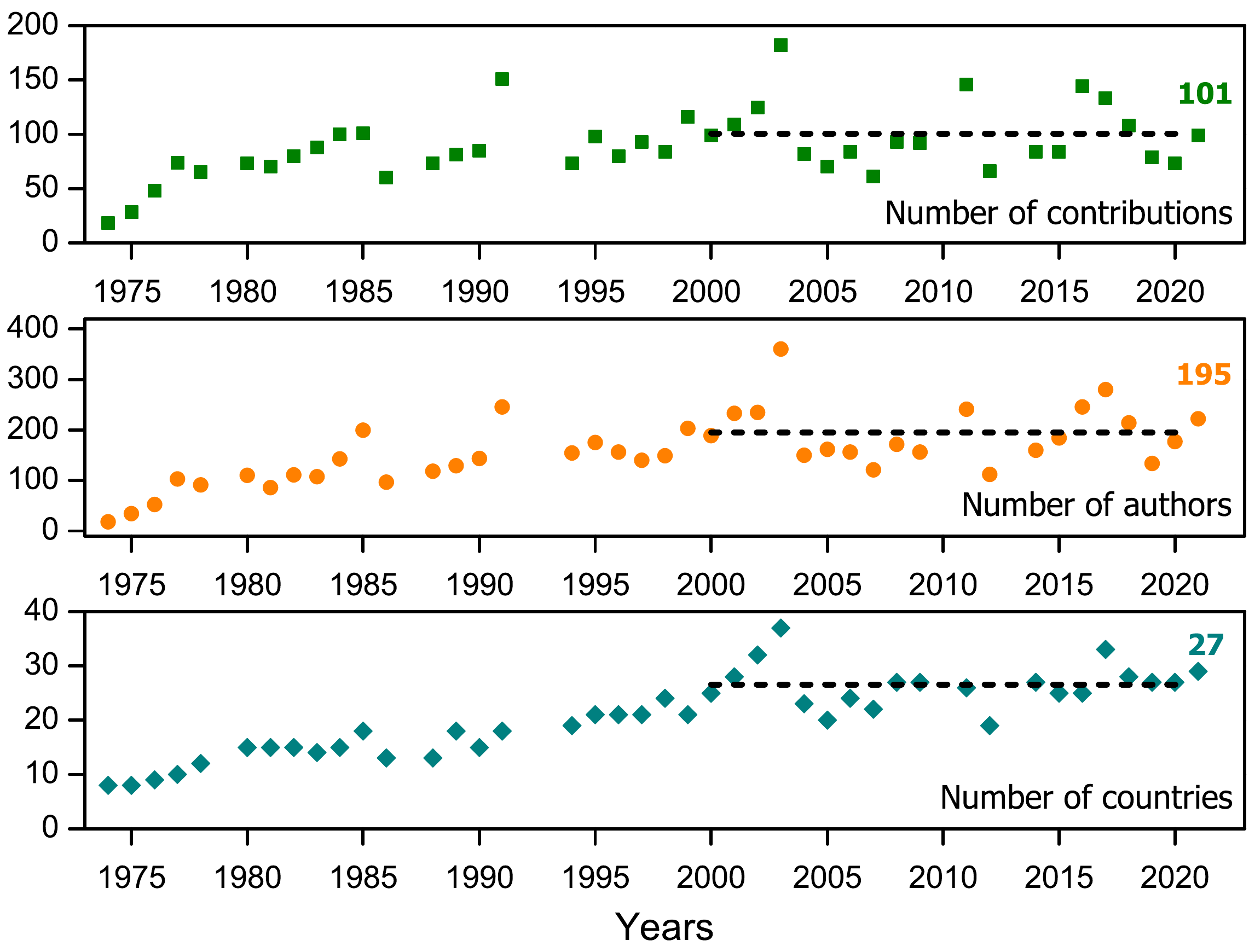}%
\hfill%
\includegraphics[width=0.48\textwidth]{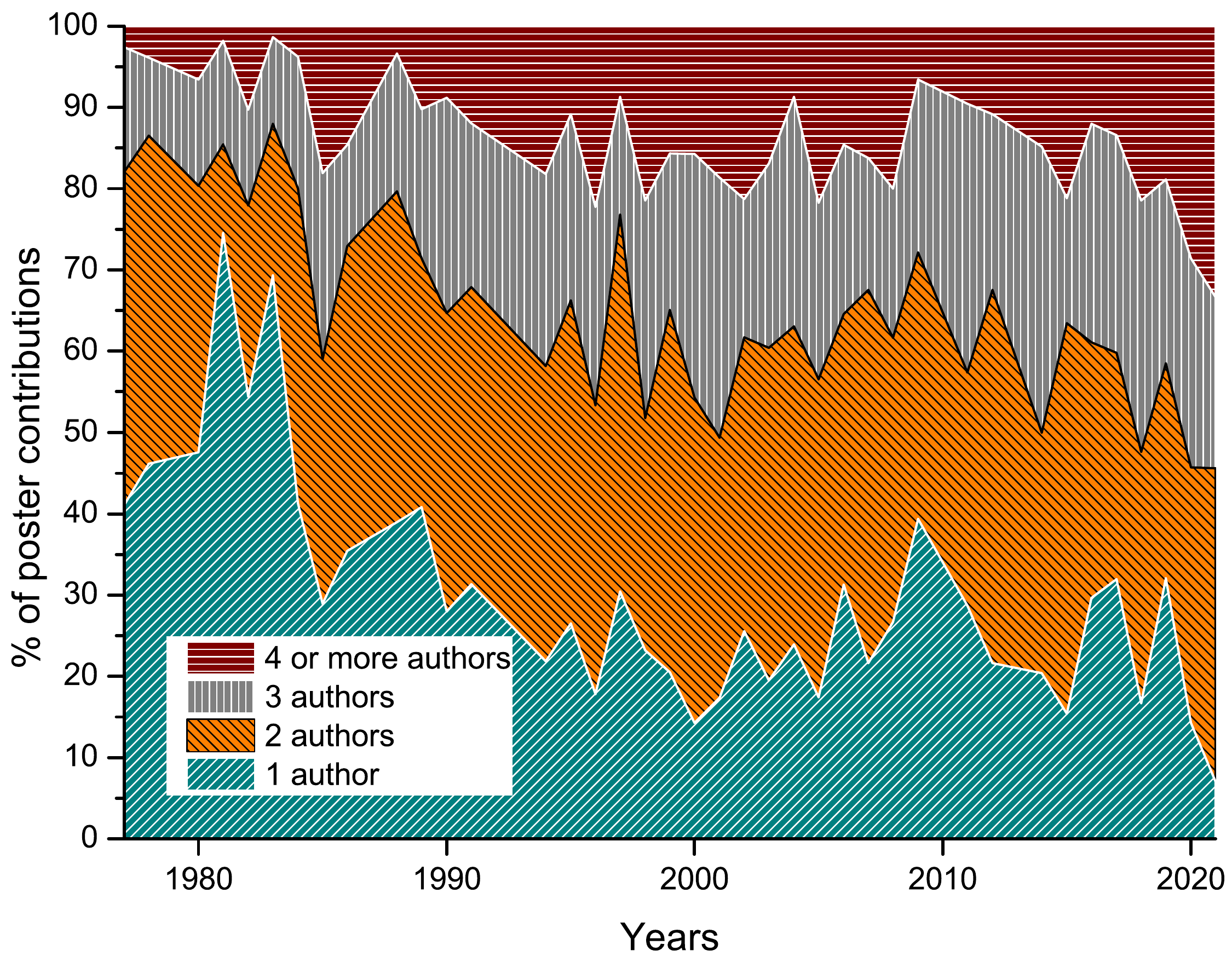}%
\\%
\parbox[t]{0.48\textwidth}{%
\caption{(Colour online) The annual numbers of contributions, unique authors and contributing countries related to MECO conferences. The broken lines indicate the corresponding mean values calculated for the period from 2000 to 2021.}
\label{Fig_annual_stat_MECO}
}%
\hfill%
\parbox[t]{0.48\textwidth}{%
\caption{(Colour online) The annual shares of MECO posters with 1, 2, 3, 4 or more authors.}
\label{Fig_num_auth_per_contr}%
}%
\end{figure}

First of all, typical questions can be asked about our data: how many authors have become part of the MECO community, how many results from statistical physics have been discussed for almost half a century. The basic statistics are as follows: 3684 unique authors from 69 countries contributed to 42 MECO meetings. In this context, we also consider the countries that are no longer on the political map: USSR, Yugoslavia, Federal Republic of Germany (FRG), German Democratic Republic (GDR), Czechoslovakia. The annual numbers of contributions, unique authors and contributing countries can be seen in figure~\ref{Fig_annual_stat_MECO}. Every year, about a hundred contributions were presented in the MECO arena. This number is naturally limited by the time frame of the conference program as well as the physical limitations of the venue. Traditionally, the MECO conference moves from one country to another, accommodating all participants in one place for these few days~\cite{Folk2021}. Thus, a saturation of the annual number of authors can be observed: usually about 195 researchers participate, and the geography of one meeting usually includes 26 countries.

Since 1977, 72\% of all contributions have been made in a form of posters, while the rest are various types of oral presentations: Keynote, Invited or Contributed Talks. This is an important change, as poster contributions are less formal and are usually appreciated by younger researchers, whereas oral presentations are usually given by more experienced colleagues. The roles of co-authors in oral talks can be specific, for example, invited lectures are more determined by the authority of the speaker and  are usually given as solo contributions.
Poster sessions are an easy way to present multiple results -- in some cases, the structure of research groups can be inferred from the poster presentations: multiple posters by one person collaborating with many others may hint at a supervisor with students, while co-authorship in different combinations is more likely to indicate on peers who work with different tasks. Unfortunately, formal verification of such speculations requires additional external data, such as individual publication histories outside MECO and affiliation records. Nevertheless, some general patterns of natural collaboration within MECO can be explored using the data related to poster contributions. The average number of authors per contribution is about 2 (more precisely, 2.3 for posters and 1.84 for various types of oral presentations), which is typical of theoretical studies~\cite{Newman2001,Newman2004}.
Figure~\ref{Fig_num_auth_per_contr} shows a trend towards a more collaborative research: e.g., the share of solo works has decreased from 48\% in 1980s (closer to 87\% for oral talks) to 22\% after 2010 (52\% for oral talks) and at the same time the share of larger groups is gradually increasing. The maximum number of co-authors per poster is 17 (12 for oral talks). However, this tendency is not a feature of statistical physics, but the general feature of research observed for different disciplines around the world \cite{Adams2018,Fortunato2018}.

It is interesting to analyse how active particular participants are. It is expected that there is a certain proportion of authors who have participated more than once. On the one hand, such members of the community can be considered the expert core, representatives of the corresponding topics of the conference. On the other hand, knowing the proportion of authors who have attended only once and those who have returned can be useful in evaluating the conferences in general. Hypothetically, these statistics might be different for predatory conferences or conferences aimed primarily at the publication of proceedings, especially if such events are occasionally organized under the auspices of the titles indexed in databases such as Scopus or Web of Sciences (these publications are considered internationally recognized and therefore desirable for authors). With rare exceptions, conference proceedings are not published for MECO \cite{Folk2021}. Moreover, its history is longer than the modern history of scientometric evaluations. Therefore, the results obtained for this study can be considered typical of the natural researchers behaviour at least in the area of statistical physics.

67.5\% of authors entered the MECO history with 1 contribution, and there are 10 authors with 30 or more contributions in total: the productivity distribution of MECO authors follows the power law $P(n) \sim n^{-1.9}$. Naturally, the most contributive authors participated in the largest number of events. However, as mentioned above, another way to contribute more is to make multiple presentations in one meeting. Indeed, about 15\% of MECO authors are those who ever had multiple contributions (more than a third of these can be considered occasional since they participated in only one event and never returned).
The typical number of multiple contributions per author per conference is 2, but sometimes one person can be found in many co-author lists in a single year. Manual inspection shows that this can happen when a research group leader gives an oral talk and his/her pupils present a number of results in a poster session. For example, such group participation is observed in 1999 (M. Schulz from the Martin Luther University Halle with 1 oral and 7 poster presentations), 2009 (W. Janke from the Leipzig University with 1 oral and 10 poster  presentations), 2012 (G. Gulpinar from the Dokuz Eylul University in Izmir with 1 oral and 8 poster presentations).

In addition to the expert core of the conference (the returning authors, the most productive and collaborative), an important part of any conference is the newcomers. They provide a flow of fresh ideas and exchange of knowledge with the external system. The participation of newcomers means that the conference is an open and evolving system, not targeted at any local audience. According to our data, 53\% of MECO authors every year are newcomers: almost half of them (47.3\%) join being already connected to the former MECO participants, 36.5\% of new contributors join collaborating with other newcomers and only 16.2\% of new authors contribute with a solo-work. After all, 27\% of all newcomers become part of MECO community by attending subsequent meetings.

All these statistics give us an impression of a balanced expert system with a stable core of actors and a variable fraction necessary for external exchange of knowledge. However, additional analysis is needed to explicitly investigate the structure of interlinks. The conference community, naturally, is connected not only by formal co-authorship links, but also by social and informal ties. Unfortunately, only documented relationships can be used for quantitative analysis, and our dataset only reflects co-authorship connections within the MECO community. A set of useful tools is provided by complex network theory, which allows us to formalize the relationships between data in the form of mathematical network consisting of nodes connected by links or edges, e.g., see \cite{Barabasi2016}. A co-authorship network is one of the classical objects for such analysis: individual authors are represented by nodes, and a link between a pair of nodes indicates the existence of  at least one common publication.

\begin{table}[ht] \caption{The numerical characteristics of
the co-authorship network for MECO conference (42 events are considered within the period 1974--2021). $N$: number of
nodes; $L$: number of links; $\langle k \rangle$,
$k_{\mathrm{max}}$: the mean and maximum node degree,
respectively; $\langle C \rangle$: the mean clustering coefficient; global transitivity $T$; $\langle l \rangle$,
$l_{\mathrm{max}}$: the mean and maximum shortest path length; $N_{\mathrm{LCC}}$, $N_{\mathrm{nLCC}}$: size of the largest (LCC) and the next-largest (nLCC) connected components, respectively; $N_{\mathrm{i}}$: number of isolated nodes.}%
\vspace{1ex}
\begin{center}
 \label{tab_Coauth_numbers}
 {\footnotesize
\begin{tabular}{|c|c|c|c|c|c|c|c|c|c|c|c|}
\hline \textbf{Parameter}&$N$&$L$&$k_{\mathrm{max}}$&$\langle
k\rangle$&$\langle C \rangle$&$T$&$\langle l\rangle$&$l_{\mathrm{max}}$&$N_{\mathrm{LCC}}$&$N_{\mathrm{nLCC}} $&$N_{\mathrm{i}}$\\\hline \hline \textbf{Value}&
3684&6031& 46 &3.27 & 0.82& 0.46&7.25& 17&1627 (44.16\%)&34 (0.92\%)& 397 (10.8\%)
\\\hline
\end{tabular}}
\end{center}
\end{table}
The constructed co-authorship network of MECO contains 3684 nodes (authors) connected by 6031 links. The corresponding values of  basic network parameters are shown in table~\ref{tab_Coauth_numbers}. The mathematics behind can be easily found in the network literature \cite{Barabasi2016,networks3}. However, the interpretation of these numbers depends on the context of data. One can see from the table~\ref{tab_Coauth_numbers} that there is at least one author who had 46 coauthors within MECO --- indeed, W. Janke (Leipzig University) collaborated with 46 other MECO authors (the next most collaborative are A. Cuccoli (University of Florence) and R. Blinc (Jo\v{z}ef Stefan Institute, Ljubljana)  who both have 41 MECO coauthors). Such authors are usually authoritative researchers with numerous disciples and the corresponding nodes --- so-called hubs --- are important for general connectivity of the entire system \cite{Barabasi2016}. Only 1\% of all authors have 20 or more collaborators while each third is represented by an isolated node or is linked only with another author. The tail of the node degree distribution can be approximated by power law  $P(k) \sim k^{-2.6}$. The average node degree which is close to 3 means that MECO author collaborates with 3 other MECO authors on average.

The further piece of results helps us to speculate about the ways of inviting new participants to the conference. 397 authors have entered the MECO community independently, i.e., with single-authored contributions. The most consistent solo-participation is found for P.~Rusek (Wroclaw University of Technology) who contributed 7 times and every time without co-authors. Almost half of all the authors (over 44\%) are interconnected by the chains of collaboration links within the largest connected component. This largest connected fragment of coauthorship network is much larger compared to the next-largest connected component which combines less than 1\% of authors --- a typical feature of many real networks~\cite{Newman2001,Barabasi2002}. During MECO co-authorship network growth, the difference between the largest and the next-largest connected component became remarkable starting from 1995 when at least 1 thousand of authors were linked by at least 1.5 thousand  links.
For a randomly chosen author in this network, it takes only 7 steps on average (the maximum of the shortest distances is 17) to reach any other author. This is the way how scientific ideas can circulate within academic community. Adding new data (i.e., new nodes and links) with every subsequent MECO meeting, this distance gradually increased and starting from 1997 fluctuated around 7.5.
According to table~\ref{tab_Coauth_numbers}, the constructed network is well-correlated locally (the probability of the nearest neighbours of nodes to be interconnected is equal to 0.82 on average); the global connectivity of the network is lower, as it is indicated by the value of transitivity.

In a similar way, the structure of relations between countries in the context of MECO can be studied. Unfortunately, only 809 (about 22\%) of MECO contributions are international, but the trend towards more collaborative research works can be seen also on this scale:  the proportion of contributions mentioning a single country has gradually decreased from over 90\% in the 1970s to $\approx 74\%$ since 1995. To give an example,  the contribution ``Superionic liquids in slit nanopores: Bethe-lattice approximation and Monte Carlo simulations'' brings together six different countries (Belarus, United Kingdom, Ukraine, Germany, Poland, France) in 2021. Based on these data, another co-authorship network is constructed: this time the nodes represent countries and the link between two of them indicates that both are mentioned in the affiliation field of the same MECO contribution. In the weighted version of this network, different coefficients are assigned to links: the strength of each is proportional to the number of common contributions of the corresponding countries.
Let us start again with the basic network parameters, see table~\ref{tab_Coauth_countr_numbers}. This network is much smaller and much more connected: almost 90\% of countries are interconnected by formal collaboration links in the context of MECO conference belonging to the single connected component and mutually reachable typically via two steps (the maximum distance is 4). 7 isolated nodes correspond to: Kazakhstan, Lithuania, Moldova, Serbia, Morocco, Colombia, Iraq. The most collaborative country with 43 international links is Germany (not taking into account the statistics for FRG and GDR) which is followed by France (41 links) and United Kingdom (31). The other countries typically have 8--9 closest neighbours in the collaboration network. Germany is also the country with the largest contribution to MECO, even if only statistics after 1990 (when GDR and FRG were united) is taken into account. It is followed by Poland, France, Hungary, Italy and Ukraine --- it is interesting that the balance between Eastern and Western Europe is reflected in this TOP6 list. The strongest links in the network connect Germany and Ukraine (39 common papers); Germany and Hungary (38); Germany and France (37); Austria and Ukraine (35); Germany and USA (34).
\begin{table}[ht] \caption{The numerical characteristics of
the co-authorship network at the level of countries for MECO conference (42 events are considered within the period 1974--2021). The notations are the same as in the table~\ref{tab_Coauth_numbers}.}%
\vspace{1ex}
\begin{center}
 \label{tab_Coauth_countr_numbers}
 {\footnotesize
\begin{tabular}{|c|c|c|c|c|c|c|c|c|c|c|}
\hline Parameter&$N$&$L$&$k_{\mathrm{max}}$&$\langle
k\rangle$&$\langle C \rangle$&$T$&$\langle l\rangle$&$l_{\mathrm{max}}$&$N_{\mathrm{LCC}}$&$N_{\mathrm{i}}$\\\hline Value&
69&300& 43 &8.7 &0.68 & 0.40 &2& 4&62 (89.9\%)& 7 (10.1\%)
\\\hline
\end{tabular}}
\end{center}
\end{table}

\begin{figure}[!t]
\centerline{\includegraphics[width=0.68\textwidth]{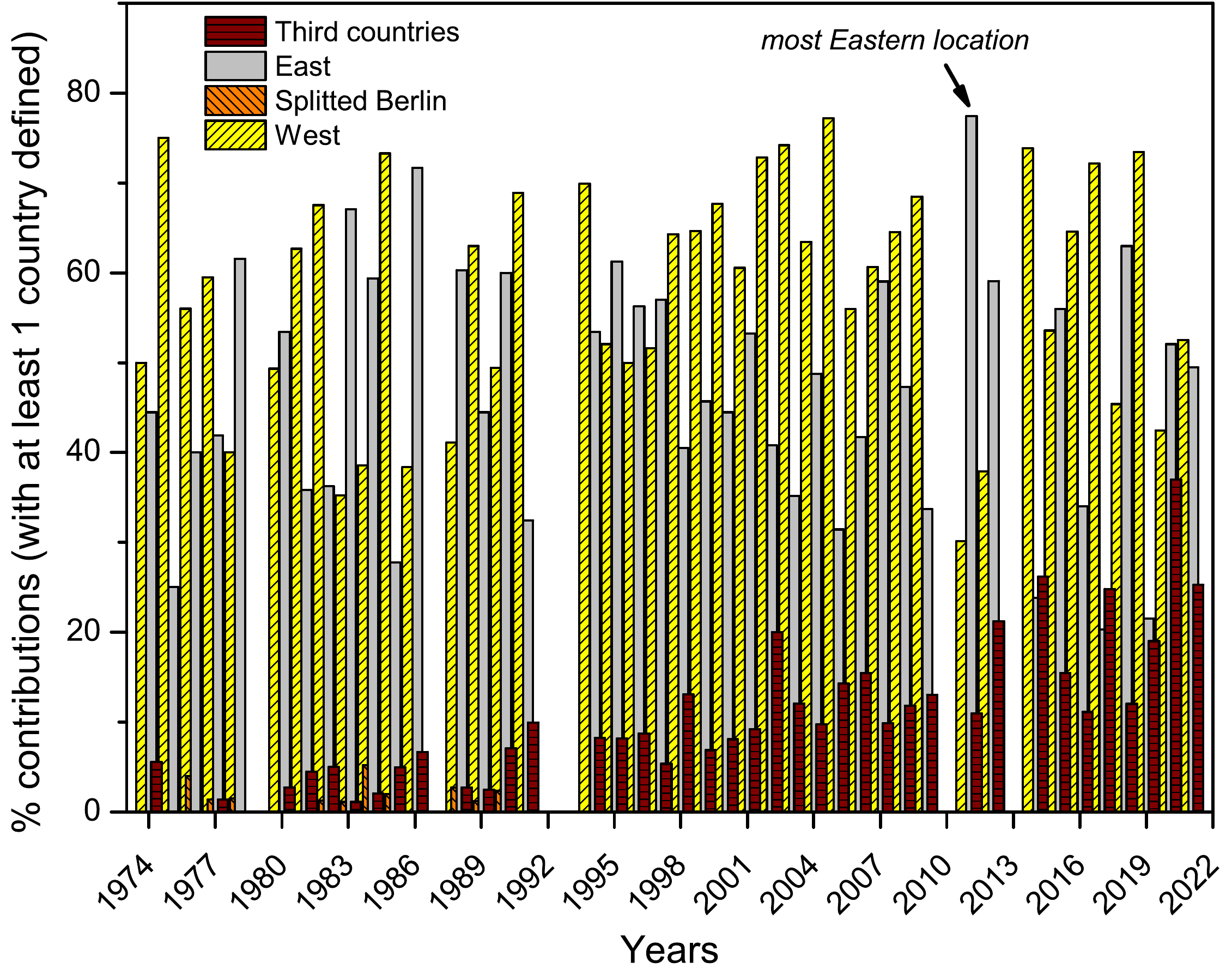}}%
\caption{(Colour online) The annual shares of MECO contributions from Eastern, Western European parts or other countries. The arrow points at the values which correspond to the year when MECO was held in its most eastern geographical point (Lviv, Ukraine). The shares of the Eastern countries are marked by stars if MECO location was on the Eastern part in the corresponding year.}
\label{Fig_East-West}%
\end{figure}

Since the idea of bringing together the physicists from East and West Europe was behind MECO from the very beginning, it is interesting to investigate the  proportions of the respective contributions. Each paper where at least one country is defined, is marked with one or more following labels: ``East'' (at least one country on the East from the former Iron Curtain is found in the affiliation list), ``West'' (at least one Western country is mentioned, correspondingly), ``Splitted Berlin'' (while the belonging of other German cities to Eastern of Western part of Europe before 1991 can be determined, it is not possible to do so for Berlin) and ``Third Countries'' (non-European countries). The proportions of contributions related to each category are shown in figure~\ref{Fig_East-West}. Here, since more than one label can be assigned to a paper, the total sum of shares can be greater than 100\%. Although the majority of contributions in most cases correspond to western countries, it can be seen that the situation is reversed for the years when MECO was located in the Eastern part of Europe. One exception is 1995, when MECO was hosted by the Institute for Theoretical Physics of the University of Linz in Puchberg/Wels (Austria). The largest share of contributions from Eastern countries was made in 2011, when MECO was held in its easternmost geographical point --- Lviv (Ukraine) --- hosted by the  Institute for Condensed Matter Physics of the National Academy of Sciences of Ukraine (ICMP).

\section{The role of Prof.~Berche in MECO and the importance of MECO for Prof.~Berche}\label{BB}

As it was mentioned above, Bertrand Berche joined to MECO community in 1991. He was initially connected with two former MECO participants: F.~Igloi (Institute for Solid State Physics and Optics, Budapest) and L.~Turban (Universit\'{e} de Nancy I). By 2021, Bertrand had already established direct collaboration links with the other 28 members of the MECO community. Moreover, 9 of them entered the MECO community for the first time, already being connected with Prof.~Berche or being part of a group of authors, where Bertrand has got the longest MECO history. In other words, one can conclude that Bertrand played an important role in bringing newcomers to MECO. One of them was W.~Janke who later became one of the key players in the MECO community as the most prolific author of more than 70 contributions and the largest number of co-authors, as it is mentioned above. In addition, Bertrand helped to enrich the geographical landscape of the MECO conference through his collaboration with authors from Venezuela. This country appeared in the list of contributive countries in 2002 for the first time.

\begin{figure}[h]
\centerline{\includegraphics[width=0.85\textwidth]{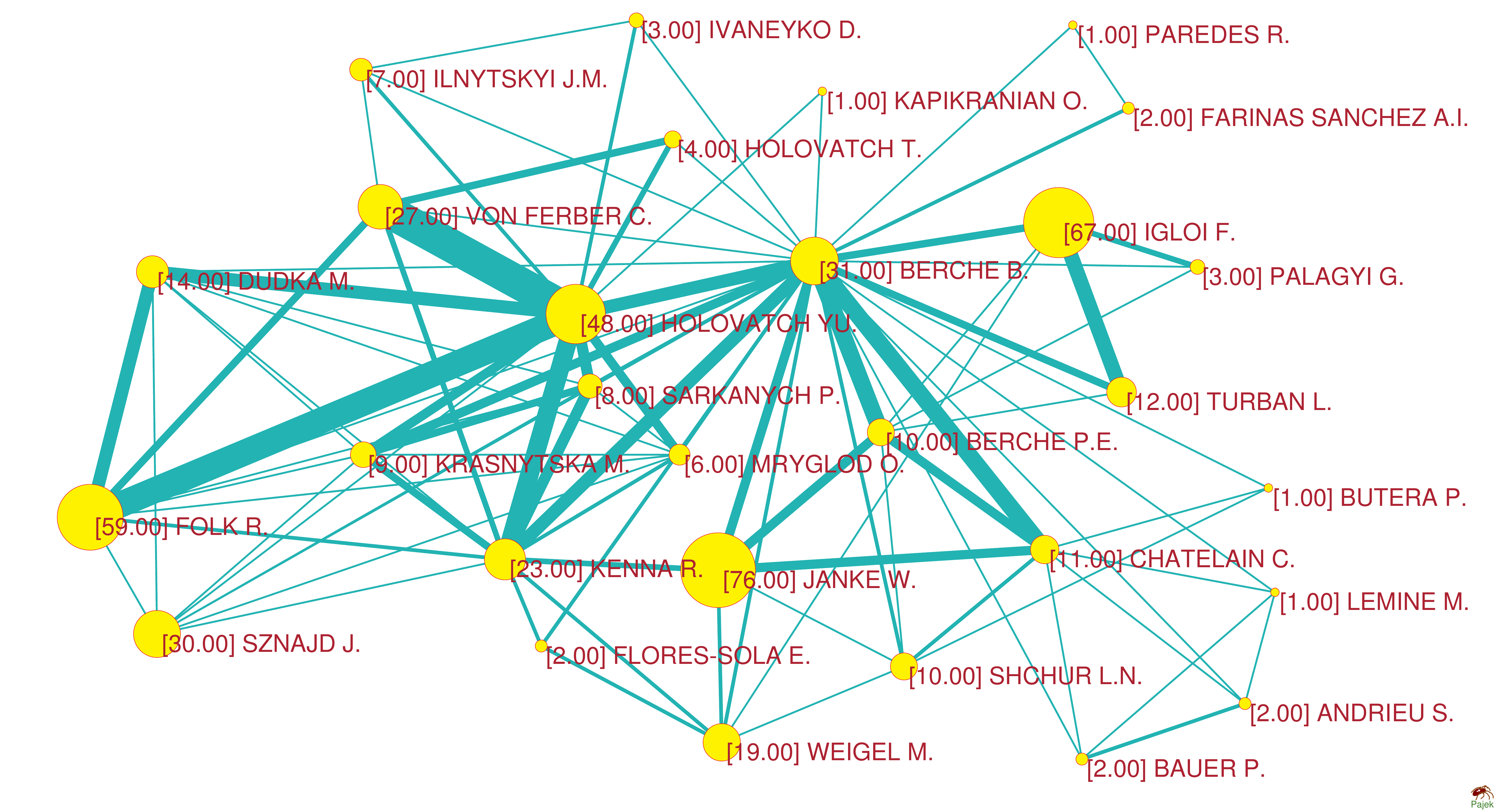}}%
\caption{(Colour online) The ego-centric fragment of co-authorship network of MECO conference in respect to the node ``Bertrand Berche'': each node represents Bertrand's MECO co-author during the observed period (the total numbers of MECO contributions are shown in the square brackets near the names and also reflected in the sizes of nodes). The weight of a link is proportional to the number of co-authored MECO contributions.}
\label{Fig_ego-coauth}%
\end{figure}
\begin{figure}[!h]
\centerline{\includegraphics[width=0.98\textwidth]{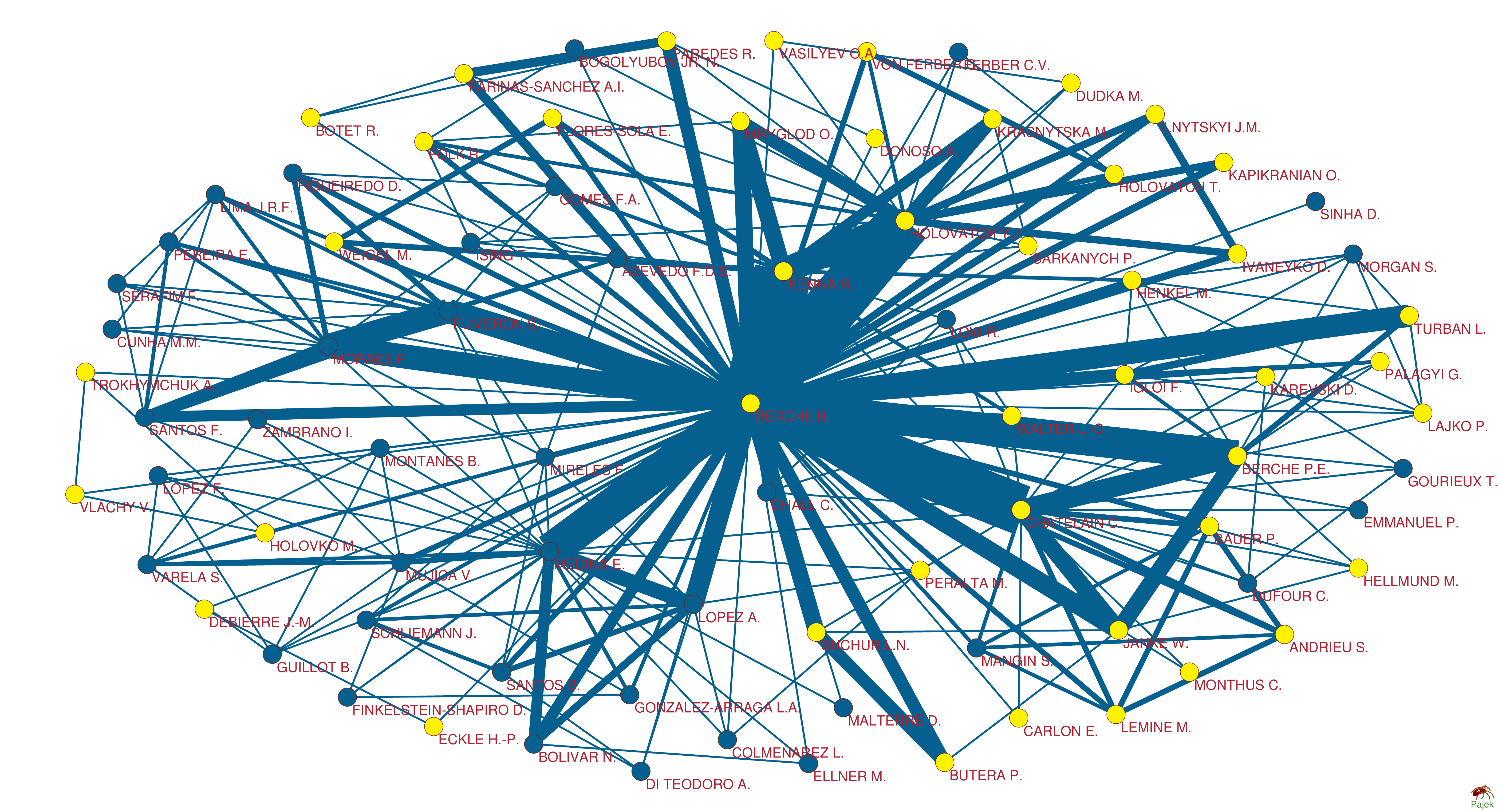}}%
\caption{(Colour online) The co-authorship network of Bertrand Berche based on the publication data [1989--2021] obtained from Scopus database. The weight of a link is proportional to the number of papers, co-authored with Bertrand. The nodes which represent the authors that ever participated in the MECO conference are marked with the lighter color (yellow online).}
\label{Fig_ego-coauth_scopus}%
\end{figure}
Bertrand Berche belongs to the largest connected component of co-authorship network of the MECO conference. He is in the TOP15 of the authors with the largest number of direct collaborators and in the TOP15 of the ones with the highest value of betweenness centrality which indicates the importance of a selected node or link for creating the shortest paths between pairs of other nodes, see, e.g., \cite{Barabasi2016}. The high value of closeness centrality (actually the second largest value in the network) demonstrates the central position of Bertrand Berche in the community, relatively close to any other participant. The ego-centric fragment of MECO co-authorship network for Bertrand Berche is shown in figure~\ref{Fig_ego-coauth}: it can be seen that each of 28 other authors is connected to Bertrand; other links within this group are shown as well. The width of links in figure~\ref{Fig_ego-coauth} is proportional to the number of common MECO contributions: the most intensive collaboration is found with C.~Chatelain (University of Lorraine, Nancy), P. E.~Berche (Universit\'{e} de Nancy I), R.~Kenna (Coventry University) and Yu.~Holovatch (ICMP, Lviv).

To investigate the hypothetical impact of MECO on the career of Bertrand Berche, his personal publication history was examined.
144 documents authored by Bertrand Berche were found in the Scopus database\footnote{\url{https://www.scopus.com}} on October 27, 2022. To compare Scopus data with our MECO dataset, let us consider the publication window up to the year 2021 inclusive (so, the three most recent papers are omitted further).  According to these data, his academic career began in 1989. Since Bertrand first participated in  MECO in 1991, one can conclude that these surroundings made an impact on his professional life almost from the very beginning. Indeed, more than half of Bertrand's 81 co-authors were part of the MECO community, see figure~\ref{Fig_ego-coauth_scopus}: 30 participated in meetings before the first joint paper with Bertrand, and 13 joined the MECO community afterwards. In addition, the majority of Bertrand Berche's 108 papers (76.6\%) are published in co-authorship with the colleagues that participated in MECO at least once. The strongest collaboration links connect Bertrand with Kenna~R. (31 joint papers); Holovatch~Yu. (29); Chatelain~C. (27); Medina~E. (19); Berche~P.E. (17) --- and almost all of them are active MECO participants as well (Medina~E. is not in our MECO dataset and thus, the corresponding node is shown in dark color in figure~\ref{Fig_ego-coauth_scopus}, although he presented a poster on MECO in 2010).

\section{Summary}\label{summary}
The approaches to a quantitative analysis of publication data described in this paper are not innovative or sophisticated, but the obtained results provide another piece of puzzle in understanding scientific conferences as  special kinds of interaction and communication between researchers. The analysis of bibliographic data is used here to demonstrate how to answer the questions about the MECO conference, i.e., core authors, collaboration patterns, geographic landscape, balance between stability and replenishment of the conference membership community. In particular, the tendency to more collaborative and more international contributions is demonstrated; the intention to bring together the physicists from the Eastern and Western countries of Europe  can be traced from the data. Typical network parameters as well as statistical properties which characterize scientific conference are derived.
The shift of perspective is used here in order to get new insights from the same data. On the one hand, a general analysis of MECO conference is performed. On the other hand, the role of a specific participant --- Bertrand Berche --- in the conference and, vice versa, the role of the conference in the individual academic career are articulated.

\emph{This paper is motivated not only by research issues but also by the author's purely personal desire to congratulate her dear friend Bertrand, to thank him for being part of a network which supports all of us and to wish him numerous successful professional collaboration links, as well as many warm and friendly personal contacts.}

\section*{Acknowledgements}
This work would be impossible without collective efforts of a group of interested researchers who themselves are tightly connected with MECO conferences: the preliminary idea to apply quantitative approaches to MECO data belongs to R.~Kenna and P.~Sarkanych; deep understanding of MECO history and collecting the originals of printed MECO programs would not be possible without personal engagement and help of R.~Folk, J.~Sznajd, B.~Berche; M.~Dudka together with L.~Didukh spent a lot of time for scanning, recognizing and preprocessing of documents; web-page with online archive of MECO programs was designed and maintained by M.~Krasnytska.
The author also expresses separate gratitute to Yu.~Holovatch for discussing and improving this paper and especially for a nice title suggested. This work was supported in part by the Ministry of Education and Science of Ukraine, the joint Ukrainian-Austrian research project No.~0122U002589. 
\emph{And finally, author would like to thank Ukrainian Armed Forces for the defence and possibility to perform research even in times of war.}

\ukrainianpart

\title{╬фшэ чр тё│ї Єр тё│ чр юфэюую: яЁю Ёюы№ ъюэЇхЁхэЎ│┐ є цшЄЄ│ фюёы│фэшър}
\author{O. ╠Ёшуыюф}

\address{▓эёЄшЄєЄ Ї│чшъш ъюэфхэёютрэшї ёшёЄхь ═рЎ│юэры№эю┐ рърфхь│┐ эрєъ ╙ъЁр┐эш, \\тєы. ╤т║эЎ│Ў№ъюую, 1, 79011 ╦№т│т, ╙ъЁр┐эр;  ╤яiтяЁрЎ  \textbf{L}$^4$ i ╩юыхфц фюъЄюЁрэЄiт ``╤ЄрЄшёЄшўэр Їiчшър ёъырфэшї ёшёЄхь'',
	╦ щяЎi┤-╦юЄрЁшэуi -╦№тiт-╩ютхэЄЁi, ктЁюяр }
\makeukrtitle

\begin{abstract}
	\tolerance=3000%
╙ ЁюсюЄ│ яЁхфёЄртыхэю ъ│ы№ъ│ёэшщ юяшё эрєъютю┐ ъюэЇхЁхэЎ│┐ MECO (Middle European Cooperation in Statistical Physics), ∙ю срчє║Є№ё  эр с│сы│юуЁрЇ│ўэшї чряшёрї. ╬суютюЁ■■Є№ё  ёЄрЄшёЄшър фюяют│фхщ Єр єўрёэшъ│т, чръюэюь│ЁэюёЄ│ ёя│тяЁрЎ│  ъ эр Ё│тэ│ ртЄюЁ│т, Єръ │ эр Ё│тэ│ ъЁр┐э, Єшяют│ яЁюяюЁЎ│┐ эютрўъ│т Єр яюёЄ│щэшї єўрёэшъ│т Єр │э°│ їрЁръЄхЁшёЄшъш эрєъютю┐ яюф│┐.
╨хчєы№ЄрЄш Ў№юую ўрёЄъютюую фюёы│фцхээ  ║ тэхёъюь фю ъЁр∙юую Ёючєь│ээ  °ы ї│т ЇюЁьры│чрЎ│┐ Єр юЎ│э■трээ  ъюэЇхЁхэЎ│щ Єр ┐ї Ёюы│ т │эфшт│фєры№эшї ърЁ'║Ёрї фюёы│фэшъ│т. ╟ ьхЄю■ фхЄры№э│°х фюёы│фшЄш юёЄрээ║, тшъюэрэю чь│∙хээ  яхЁёяхъЄштш: эр фюфрўє фю чруры№эюую рэры│чє ъюэЇхЁхэЎ│щэшї фрэшї, чрёЄюёютє║Є№ё  хуюЎхэЄЁшўэшщ я│фї│ф фы  тшф│ыхээ  Ёюы│ юъЁхьюую єўрёэшър є ъюэЇхЁхэЎ│┐ Єр, эртяръш, Ёюы│ ╠┼╤╬ є яЁюЇхё│щэюьє цшЄЄ│ тўхэюую. ╓  ёЄрЄЄ  ║ ўрёЄшэю■ ёяхЎ│ры№эюую тшяєёъє ╤╠╨, ∙ю яЁшєЁюўхэшщ ■т│ых■ ┴хЁЄЁрэр ┴хЁ°р --- чэрэюую Ї│чшър, ръЄштэюую ўыхэр ёя│ы№эюЄш ртЄюЁ│т Єр ЁхфръЄюЁ│т цєЁэрыє, фртэ№юую ёя│тртЄюЁр Єр фюЁюуюую фЁєур ртЄюЁъш.

	
\end{abstract}

\end{document}